\documentclass[runningheads]{llncs}
\usepackage[T1]{fontenc}
\usepackage{graphicx,verbatim}
\usepackage[misc]{ifsym}
\usepackage{amsmath}
\usepackage{amsfonts}
\usepackage{xcolor}
\newcommand{\myvec}[1]{\overset{\scriptscriptstyle\rightarrow}{#1}}

\begin{document}
\title{$Q$-space Guided Collaborative Attention Translation Network for Flexible Diffusion-Weighted Images Synthesis}
\titlerunning{Q-CATN}

\author{Pengli Zhu$^{1}$
\and Yingji Fu$^{1}$
\and Nanguang Chen$^{2}$
\and Anqi Qiu$^{1,2,3}$\textsuperscript{(\Letter)}
}

\institute{
$^{1}$ Department of Health Technology and Informatics, Hong Kong Polytechnic University, Hong Kong\\
\email{an-qi.qiu@polyu.edu.hk} \\
$^{2}$ Department of Biomedical Engineering, National University of Singapore, Singapore\\
$^{3}$ Department of Biomedical Engineering, The Johns Hopkins University, USA 
}

\authorrunning{P. Zhu et al.}


\maketitle              
\vspace{-10pt}
\begin{abstract}
This study, we propose a novel $Q$-space Guided Collaborative Attention Translation Networks (\textbf{Q-CATN}) for multi-shell, high-angular resolution DWI (MS-HARDI) synthesis from \textit{flexible} $q$-space sampling, leveraging the commonly acquired structural MRI data. Q-CATN employs a \textit{collaborative attention mechanism} to effectively extract complementary information from multiple modalities and \textit{dynamically} adjust its internal representations based on flexible $q$-space information, eliminating the need for fixed sampling schemes. Additionally, we introduce a range of task-specific constraints to preserve \textit{anatomical fidelity} in DWI, enabling Q-CATN to accurately learn the intrinsic relationships between directional DWI signal distributions and $q$-space. Extensive experiments on the Human Connectome Project (HCP) dataset demonstrate that Q-CATN outperforms existing methods, including 1D-qDL, 2D-qDL, MESC-SD, and QGAN, in estimating parameter maps and fiber tracts both quantitatively and qualitatively, while preserving fine-grained details. Notably, its ability to accommodate \textit{flexible $q$-space sampling} highlights its potential as a promising toolkit for clinical and research applications. Our code is available at \url{https://github.com/Idea89560041/Q-CATN}.

\keywords{DWI Synthesis  \and Conditional Generative Model \and Collaborative Attention Translation.}

\end{abstract}

\section{Introduction}
Diffusion-weighted imaging (DWI) is a key non-invasive method for evaluating brain microstructure and connectivity, providing critical insights into development, aging, and neurodegenerative diseases. Advanced models like neurite orientation dispersion and density imaging (NODDI) \cite{zhang2012noddi} and diffusion kurtosis imaging (DKI) \cite{jensen2005diffusional} offer superior tissue microstructure analysis over conventional diffusion tensor imaging (DTI). However, these methods require extensive $q$-space sampling, longer acquisition times, and complex computational processes, increasing susceptibility to motion artifacts, eddy current distortions, and physiological noise, which can affect their quantitative precision.

Recent advances in deep learning have shown promise in medical image synthesis, with several studies \cite{chen2023deep,park2021diffnet,yang2023towards} estimating parameter maps from DWI using limited gradient directions. $Q$-space deep learning (qDL) \cite{golkov2016q} pioneered the direct mapping of sparsely sampled $q$-space DWI signals to microstructural parameters via a multilayer perceptron. 
Subsequent enhancements integrated 2D spatial information \cite{gibbons2019simultaneous} and 3D sparse spatial patch representations with modified LSTM networks \cite{ye2020improved}. However, these methods are limited to generating fixed parameter maps, restricting their applicability to variably sampled DWI data. Thus, a more flexible DWI synthesis approach is needed to broaden its practical utility.

According to the principles of DWI, its generation involves a complex nonlinear relationship in $q$-space \cite{mori2007introduction}, making conditional generative adversarial networks (cGANs) \cite{isola2017image,karras2019style} well-suited for DWI synthesis. Recent studies have shown their effectiveness in producing high-fidelity medical images, such as translating structural/functional MRI into DWI \cite{tian2020deepdti} and generating DWI-derived scalar maps \cite{gu2019generating}. Complementary modalities like T1- and T2-weighted MRI have also been shown to improve DWI synthesis \cite{ren2021q,cicimen2024image}. However, existing methods are typically optimized for fixed $q$-space sampling aligned with their training datasets, limiting their applicability in clinical settings where heterogeneous sampling is common and site-specific data is often insufficient for training models. This highlights the need for a flexible DWI synthesis approach not constrained by predefined sampling strategies. While $q$-space cGANs such as Q-GAN \cite{ren2021q} and aqDL \cite{zong2024attention} offer promising solutions by enabling DWI generation at arbitrary $q$-space points, they often rely on oversimplified input representations or fail to effectively capture inter-modality correlations, limiting their ability to fully utilize complementary information and constraining their generative performance.

To overcome the above limitations, we propose a novel $q$-space guided collaborative attention translation networks (Q-CATN), for multi-shell, high-angular resolution DWI (MS-HARDI) synthesis with flexible $q$-space sampling using commonly acquired structural MRI. 
The main features are outlined as follows:
\begin{enumerate}
	\item The proposed Q-CATN framework supports MS-HARDI synthesis by incorporating \textit{flexible} $q$-space conditional information, overcoming the limitations of fixed sampling strategies commonly seen in existing approaches; 
	
	\item By introducing a \textit{collaborative attention mechanism}, Q-CATN effectively extracts compatible information from single modality and complementary information across multi-modal inputs (e.g., b0, T1- and T2-weighted images), enhancing synthesis accuracy and robustness;
	
	\item The proposed framework enables the generation of \textit{densely} sampled $q$-space data, facilitating the reconstruction of various diffusion models, which significantly benefit downstream applications.
\end{enumerate}

\section{Methodology}\label{Methodology}
\subsection{Overall Architecture}
\begin{figure}[h]
	\begin{center}
		\vspace{-15pt}
		\includegraphics[width=1\textwidth]{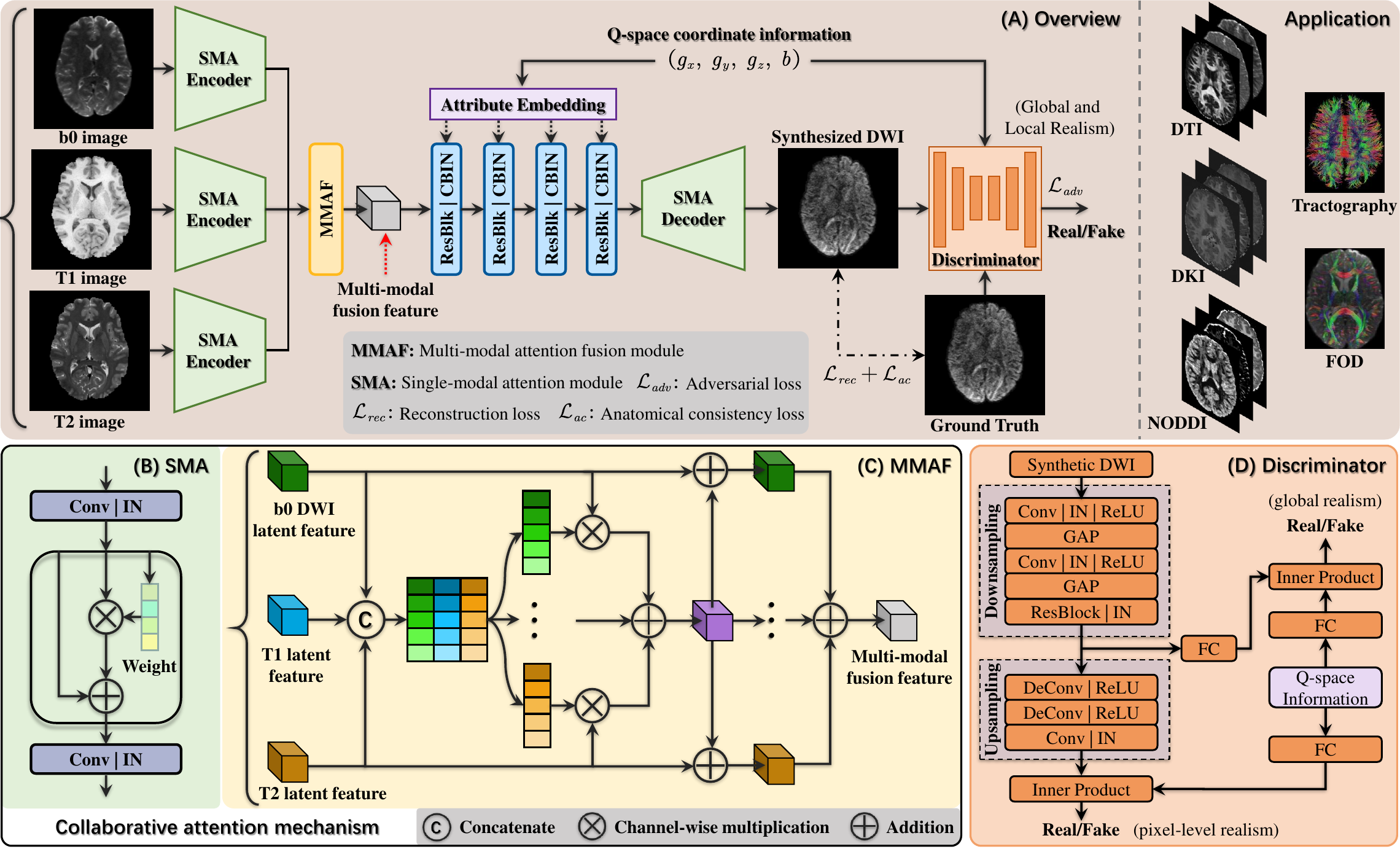}
		\vspace{-20pt}
		\caption{Overview of Q-CATN. Panel (A) illustrates the high-level structure of the model. Panel (B) details the architecture of the single-modal attention module. Panel (C) shows the multi-modal attention fusion mechanism. Panel (D) outlines the structure of the conditional discriminator.}  
		\label{fig:overall} 
	\end{center}
	\vspace{-35pt}
\end{figure}

Fig.~\ref{fig:overall} presents the Q-CATN framework, comprising a single-modal attention (SMA) encoder, a multi-modal attention fusion (MMAF) module, a $q$-space embedding module, a SMA decoder, and a conditional discriminator. Q-CATN takes $\mathbf{x}_{b0}$, $\mathbf{x}_{t1}$, and $\mathbf{x}_{t2}$ as input, representing b0, T1-, and T2-weighted images, respectively. These are processed by SMA encoders to extract latent-space features $(\mathbf{z}_{b0}, \mathbf{z}_{t1}, \mathbf{z}_{t2})$. The MMAF module integrates these features into unified representation $\mathbf{z}$, conditioned on $q$-space coordinates $\myvec{\boldsymbol{q}} = (g_x, g_y, g_z, b)$, which define the $b$-vector and $b$-value. The $q$-space embedding module plays a crucial role in transforming $\mathbf{z}$ into $\mathbf{z}_{bn}$ using $\myvec{\boldsymbol{q}}$, facilitating flexible $q$-space-aware representation learning. The SMA decoder reconstructs $\mathbf{z}_{bn}$ into the image domain as $\mathbf{x}_{bn}$ under ground-truth supervision. Meanwhile, a discriminator is trained to distinguish synthesized from real DWIs. 
Further details are provided below.
\subsection{Collaborative Attention Mechanism}
To enable the generation of precise and realistic DWI outputs, the proposed Q-CATN model employs a collaborative attention mechanism to effectively extract and integrate compatible-complementary information from multiple modalities. This architecture comprises two key components:

\vspace{5pt}
\noindent {\bf SMA Encoder}~
To improve multi-modal MRI synthesis, it is essential for each modality to provide \textit{sufficient and complementary information}. We introduce a SMA encoder to extract more compatible features $\mathbf{z}_{n}$ across three encoder branches, computed as:
	$\mathbf{z}_{n} = G_n^{enc}(\mathbf{x}_{n}) 
	+ G_n^{enc}(\mathbf{x}_{n}) \otimes \delta \left( G_n^{enc}(\mathbf{x}_{n}) \right)$,
where $G_n^{enc}(\cdot)$ denotes feature maps from the $n$-th SMA encoder, $\otimes$ represents channel-wise multiplication between feature maps and vector, and $\delta (\cdot)$ is an attention metric for DWI modality information. The SMA mechanism assigns a weight vector to each modality, emphasizing specific channels. These weighted features are combined with the input to yield an enhanced single-modal representation $\mathbf{z}_{n}$: \{$\mathbf{z}_{b0}, \mathbf{z}_{t1}, \mathbf{z}_{t2}$\}. Here, $G_n^{enc}(\cdot)$ employs strided convolutions for downsampling, while $\delta(\cdot)$ is implemented via global average pooling followed by two fully connected (FC) layers with ReLU and Sigmoid activations, respectively.

\vspace{5pt}
\noindent {\bf MMAF Module}~
To effectively integrate \textit{DWI-specific information} from multiple modalities within the collaborative attention mechanism, we introduce the MMAF module, as depicted in Fig.~\ref{fig:overall}(C). Specifically, the input modality features are concatenated and processed through FC layers to compute the attention matrix $\mathcal{A}$. 
This matrix is then used to weight the features via channel-wise multiplication. The resulting weighted features are combined to form a multi-modal feature, which is added to each input feature to produce modality-specific features $\mathbf{z}_{n}^{\text{att}}$. The multi-modal attention determines the weights $\mathcal{A}$ for the cross-modal feature channels as follows:
	$\mathcal{A} = G^{m} \left( \psi \left( \eta \left( \left( \mathbf{z}_{b0}, \mathbf{z}_{t1}, \mathbf{z}_{t2} \right) \omega_1 \right) \omega_2 \right) \right)$,
where $\omega_1$ and $\omega_2$ are mapping matrices implemented by two FC layers, while $\eta$ and $\psi$ denote ReLU and Sigmoid activations, respectively. The function $G^{m}$ applies softmax across each row of $\mathcal{A}$, ensuring each row represents modality weights for a specific pattern, and each column represents a modality's weights across all patterns. The matrix $\mathcal{A}$ is then used to extract correlated information from the multi-modal data, yielding the complementary feature $\mathbf{z}_{n}^{\text{att}}$ for the $n$-th modality:
	$\mathbf{z}_{n}^{\text{att}} = \mathbf{z}_n + \sum_{n=1}^{3} \left(\mathbf{z}_{n} \otimes \mathcal{A}_n \right)$,
where $\mathcal{A}_n$ denotes the $n$-th column of $\mathcal{A}$. The final fused feature $\mathbf{z}$ is obtained by combining these modality-specific features. 
\subsection{$Q$-space Embedding Module}
To address the limitations imposed by \textit{predefined sampling strategies}, we propose a \textit{flexible} DWI synthesis approach enabling the generation of $\mathbf{z}_{bn}$ for any given variable $\myvec{\boldsymbol{q}}$. Inspired by affine transformation parameters in normalization layers to encode attributes, we introduce central biasing instance normalization (CBIN) \cite{yu2018multi} into the residual blocks, which are dynamically modulated by the $q$-space coordinates. The operation is defined as follows:
	$\operatorname{CBIN}\left(\mathbf{z}, {\hat{\boldsymbol{q}}}\right)=\frac{\mathbf{z}-\mu\left(\mathbf{z} \right)}{\sigma\left(\mathbf{z}\right)}+b_r({\hat{\boldsymbol{q}}})$,
where $\mathbf{z}$ denotes the feature map from the previous convolution, ${\hat{\boldsymbol{q}}}$ is the $q$-space embedding code, $\mu$ and $\sigma$ are the instance mean and standard deviation, and $b_r$ is the bias for the $r$-th feature map. Finally, the SMA decoder is used to precisely reconstruct $\mathbf{z}_{bn}$ back to the direction-specific image $\mathbf{x}_{bn}$.

\subsection{$Q$-space Conditional Discriminator}
To improve the \textit{realism} of the synthesized DWI, we employ a $q$-space conditional discriminator with two levels: its bottleneck layer assesses global image realism, while the output layer evaluates pixel-level fidelity, as illustrated in Fig.~\ref{fig:overall}(D). Specifically, the discriminator processes $\mathbf{x}_{bn}$, extracting a global representation via its encoding path to evaluate global realism. Meanwhile, a decoder expands the output to match the input size, enabling per-pixel realism feedback. The $q$-space coordinates are integrated via conditional projection before the final layers of both branches. The final layer is defined as $f(\mathbf{x}_d, \myvec{\boldsymbol{q}}) := (\myvec{\boldsymbol{q}})^T V \gamma(\mathbf{x}_d) + \xi(\gamma(\mathbf{x}_d))$, where $V$ denotes a learnable embedding of $\myvec{\boldsymbol{q}}$, $\gamma(\mathbf{x}_d)$ represents the output prior to conditioning, and $\xi(\cdot)$ is a scalar function applied to $\gamma(\mathbf{x}_d)$.
\subsection{Loss Function}
\noindent {\bf Adversarial Loss}~
To enhance the realism of the synthetic DWI both \textit{locally and globally}, we introduce adversarial losses for the encoder and decoder to achieve superior outcomes. The adversarial loss $\mathcal{L}_{adv}^{(*)}$ is defined as:
\begin{equation}\label{eq:6}
	\mathcal{L}_{adv}^{(*)} = \mathbb{E}\left[\log \left(1 - D_{(*)}(\mathbf{y}_{bn}, \myvec{\boldsymbol{q}})\right)\right] + \mathbb{E}\left[\log D_{(*)}(\mathbf{x}_{bn}, \myvec{\boldsymbol{q}})\right],
\end{equation}
where ${(*)}$ denotes either the encoder and decoder of the discriminator, respectively. $\mathbf{x}_{bn}$ is the synthesized DWI conditioning on the variable $\myvec{\boldsymbol{q}}$, and $\mathbf{y}_{bn}$ represents the real tuple with $\myvec{\boldsymbol{q}}$ sampled from the training data.

\noindent {\bf Reconstruction Loss}~
To better align low-frequency details and ensure consistency with the input, we further introduce a reconstruction loss $\mathcal{L}_{rec}$ as below:
\begin{equation}\label{eq:7}
	\mathcal{L}_{rec} = 
	\begin{cases} 
		\mathbb{E} \left[ \| \mathbf{x}_{bn} - \mathbf{x}_{bn}^{*} \|_1 \right], & \text{if } b > 0 \\
		\mathbb{E} \left[ \| \mathbf{x}_{bn} - \mathbf{x}_{b0} \|_1 \right], & \text{if } b = 0 
	\end{cases}
\end{equation}
where $\mathbf{x}_{bn}^{*}$ is the reference DWI with targeted $q$-space coordinates $\myvec{\boldsymbol{q}}$.
 
\noindent {\bf Anatomical Consistency Loss}~
To ensure the synthesized DWI accurately reflects the \textit{underlying tissue microstructure}, we employ a spatially-correlative loss \cite{zheng2021spatially} to preserve image structure. The anatomical features of DWI are extracted using a VGG16 network, and self-similarity is computed as a map:
	${{\cal L}_{ac}} = \left\lVert G_{ac}\left( \mathbf{x}_{bn}\right) - G_{ac}\left(\mathbf{x}_{bn}^{*} \right) \right\rVert_1$,
where $G_{ac}({x_i})=(f_{x_i})^T(f_{x_*})$, $f_{x_i}^T$ represents the feature of a query point $x_i$, $f_{x_*}$ denotes the features associated with patch, while $G_{ac}({x_i})$ captures the anatomical correlation between different query points. 

\noindent {\bf Overall Loss Function}~
To summarize, the overall loss function for Q-CATN can be expressed as:
	${{\cal L} =  {\cal L}_{adv} + {\lambda _{rec}}{{\cal L}_{rec}} + {\lambda _{ac}}{{\cal L}_{ac}},}$
where ${\lambda _{rec}}$ and ${\lambda _{ac}}$ are loss weights used to adjust the relative importance of each term.
\subsection{Implementation}\label{sec:Implementation}
\noindent {\bf Dataset}~
We conducted experiments using preprocessed data from the Human Connectome Project (HCP) release\footnote{\url{https://www.humanconnectome.org/}}\cite{van2013wu}.  
Skull stripping was performed using brain masks, and T1- and T2-weighted images were resampled to b0 resolution, followed by image registration with DPABI \cite{yan2016dpabi}. The training set included 20 subjects (5400 DWIs), while 30 subjects were reserved for testing. Training data comprised 2D axial slices sampled from random directions, with DWI intensities normalized using the corresponding b0 and b-values scaled by their maximum.

\noindent {\bf Training and Inference}~
Q-CATN is trained using b0, T1, and T2 images as inputs, with diverse $q$-space coordinates integrated into latent features. The model generates the target DWI, while a discriminator enhances realism by distinguishing synthetic from real DWI. Optimization is performed using loss functions until convergence. During the inference, b0, T1-, T2-weighted images, and \textit{flexible $q$-space conditions} are used to synthesize MS-HARDI.

\noindent \textbf{Optimization}~
We implemented Q-CATN using Python 3.8 and PyTorch 1.13.0. During the training, we set the loss weights $\lambda_{rec}$ and $\lambda_{ac}$ to 100 and updated the discriminator every two generator updates.  
The model was trained on four NVIDIA A100 GPUs using the ADAM optimizer with a mini-batch size of 128. The initial learning rates for the generator and discriminator were set to $1 \times 10^{-4}$ and $5 \times 10^{-5}$, respectively, and reduced by a factor of 0.95 after each epoch. The training process was conducted for a maximum of 300 epochs.

\section{Experimental Results}\label{Study}
To evaluate the performance of Q-CATN, we conducted comparative experiments with benchmark models. Specifically, 1D-qDL \cite{golkov2016q}, 2D-qDL \cite{gibbons2019simultaneous}, and MESC-SD \cite{ye2020improved} were limited to generating specific parameter maps from undersampled DWI data (using 30 fixed-direction data by default), while QGAN \cite{ren2021q} and Q-CATN employed the \texttt{sphere2cart} function from DIPY \cite{garyfallidis2014dipy} to \textit{flexibly simulate dense $q$-space sampling} (270 directional data) for downstream evaluation. Ground truth in all experiments was derived from the complete testing subject.

\subsection{Qualitative Analysis}
\begin{figure}[!t]
	\centering
	\begin{minipage}{0.65\textwidth}
		\includegraphics[width=0.98\textwidth]{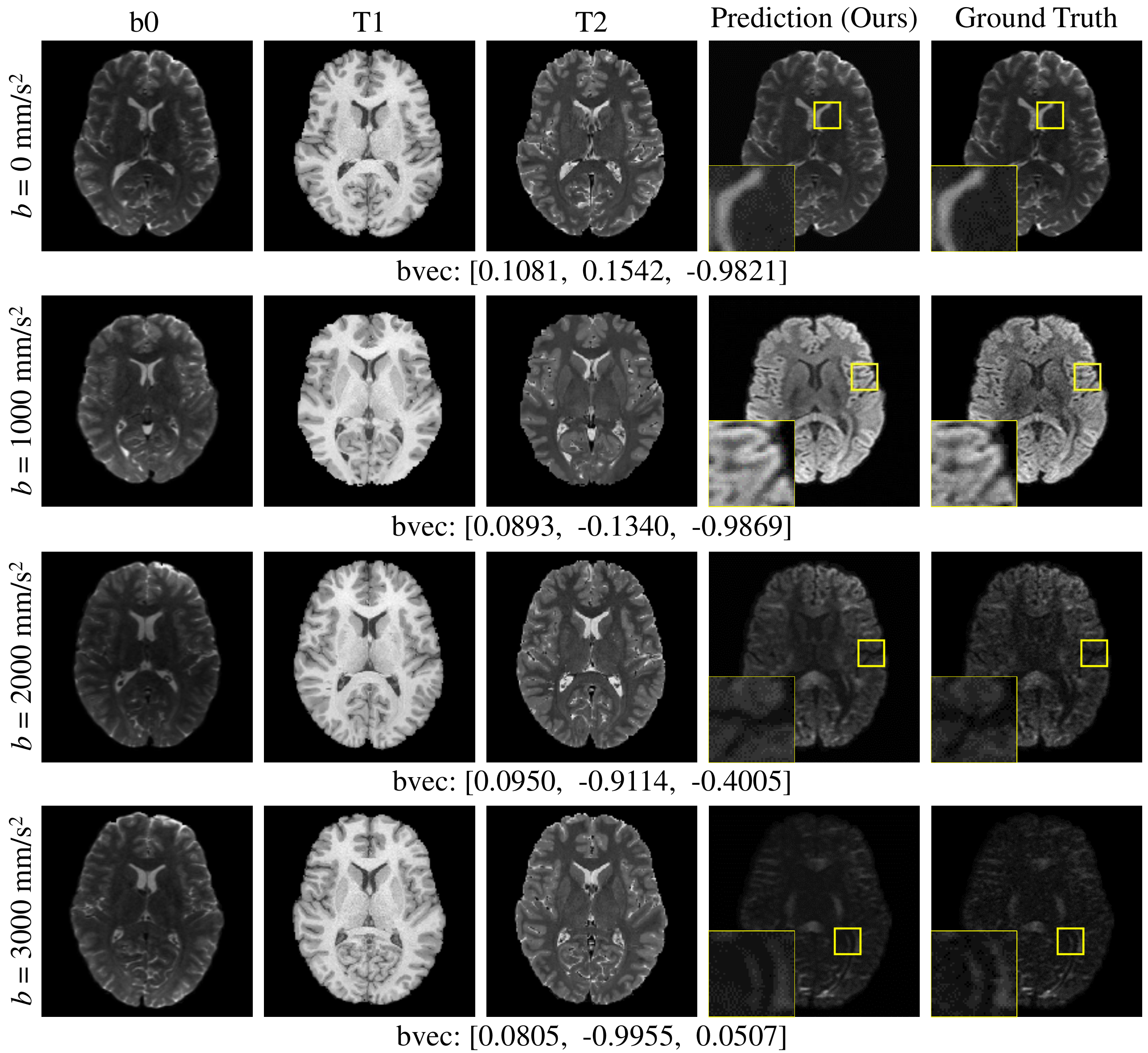}
	\end{minipage}
	\begin{minipage}{0.3\textwidth}
		\caption{DWI synthesis results under different b-values configurations. On the left, all potential input channels (b0, T1, T2) are shown, while the right side displays the predicted results, with a standard DWI slice serving as the reference.} 
		\label{fig:qualitative}
	\end{minipage}
\vspace{-6pt}
\end{figure}

\begin{figure}[!t]
	\centering
	\begin{minipage}{0.66\textwidth}
		\includegraphics[width=0.98\textwidth]{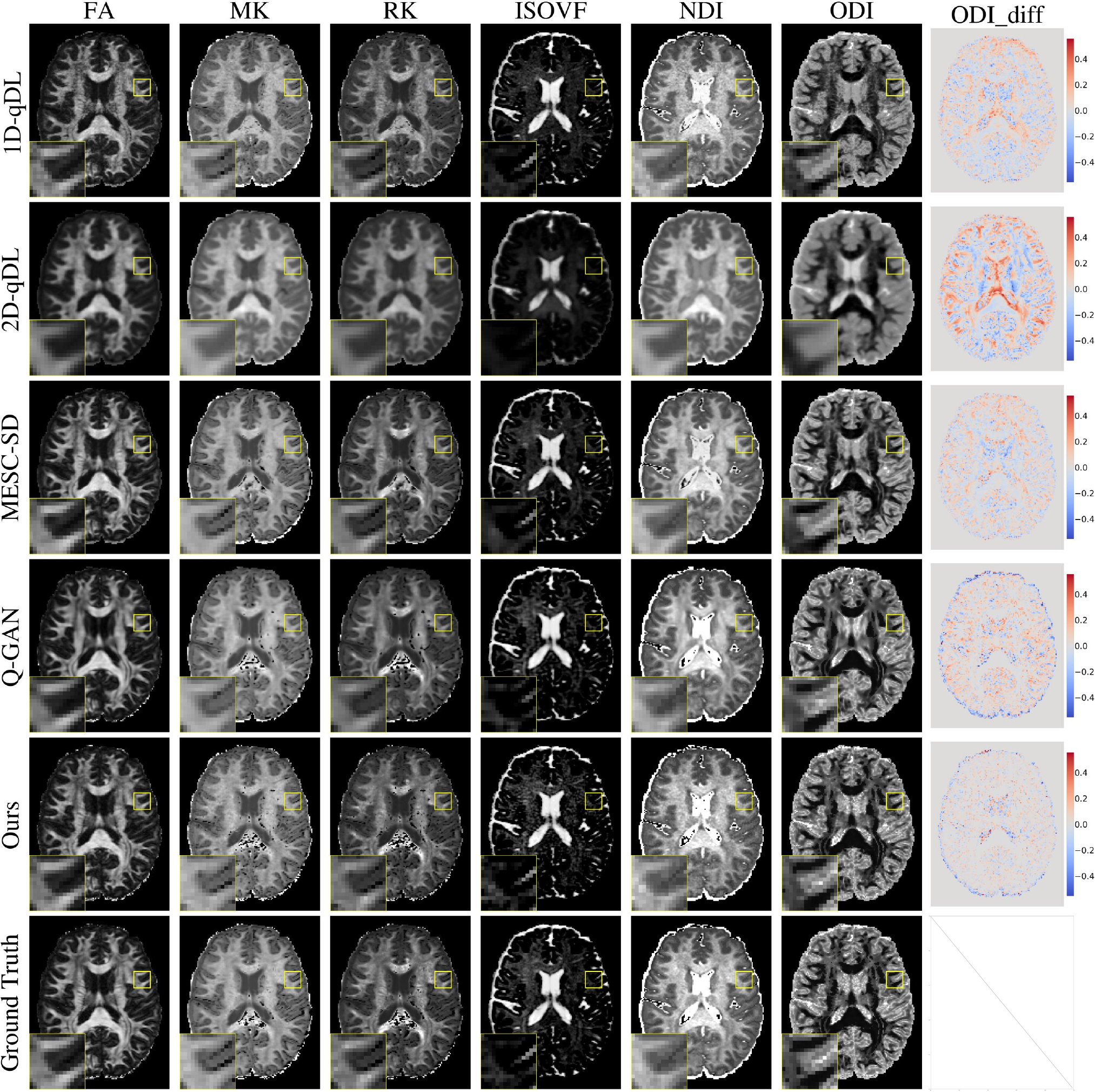}
	\end{minipage}
	\begin{minipage}{0.3\textwidth}
		\caption{Qualitative comparison of different methods for undersampling DWI parameter fitting. The first six columns show the various diffusion maps and the last column is the ODI error map, each row represents different comparison methods, with the reference map at the bottom.} 
		\label{fig:comparsion}
	\end{minipage}
	\vspace{-5pt}
\end{figure}

We conducted DWI synthesis experiments across varying b-values, as illustrated in Fig.~\ref{fig:qualitative}. From the zoom-in perspective, Q-CATN preserves fine anatomical structures, yielding results consistent with the ground-truth. 
We further evaluated against comparison methods for estimating parameter maps, performing diffusion imaging using DIPY \cite{garyfallidis2014dipy} for DTI (using only b=1000 s/mm\(^2\)) and DKI, while fitting the NODDI model using AMICO \cite{daducci2015accelerated}.
As shown in Fig.~\ref{fig:comparsion}, 1D-qDL fails to yield feasible results due to its reliance on predefined downsampling scheme and 2D-qDL exhibited notable inaccuracies in structural detail prediction. Although MESC-SD and Q-GAN demonstrated improved performance, anatomical details remained unclear upon closer inspection. In contrast, Q-CATN generated dense DWIs, producing parameter maps that closely align with the ground truth, surpassing all other methods.

\subsection{Quantitative Analysis}
\begin{figure}[!t]
	\begin{center}
		\includegraphics[width=0.94\textwidth]{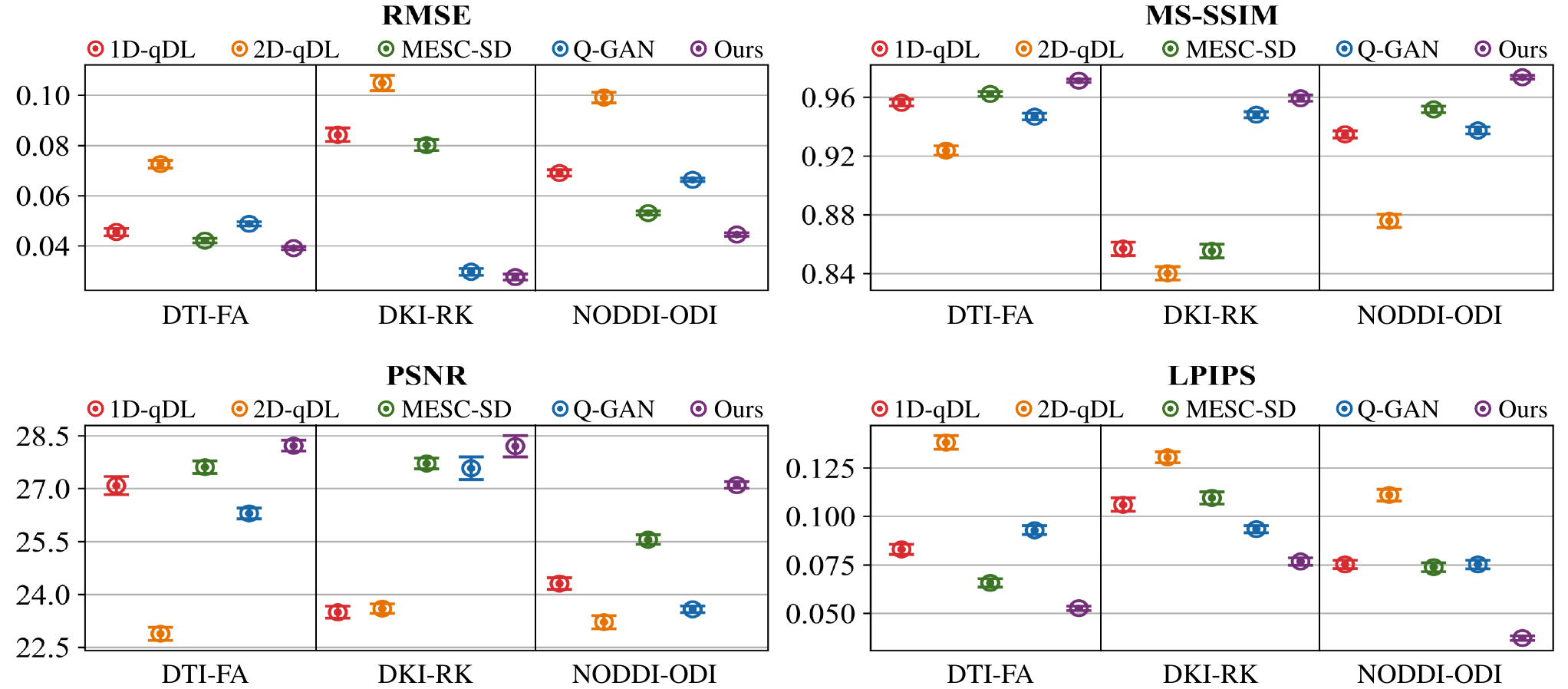}
		\vspace{-12pt}
		\caption{Quantitative comparison of estimated parameter maps in different methods.} 
		\label{fig:quantitate}
	\end{center}
	\vspace{-15pt}
\end{figure}
To quantitatively assess the effectiveness in parameter fitting, we utilized several image quality metrics, including root mean square error (RMSE), multi-scale structural similarity index (MS-SSIM), peak signal-to-noise ratio (PSNR) and Learned Perceptual Image Patch Similarity (LPIPS). Except for RMSE and LPIPS, higher values denote better performance. Fig.~\ref{fig:quantitate} summarizes the quantitative comparison of estimated diffusion maps, Q-CATN consistently surpasses all baseline models across four metrics, demonstrating its superior ability to adapt to proposed collaborative
attention and flexible $q$-space sampling schemes.
\subsection{FOD and Tractography}
\begin{figure}[!t]
	\centering
	\begin{minipage}{0.65\textwidth}
		\includegraphics[width=0.98\textwidth]{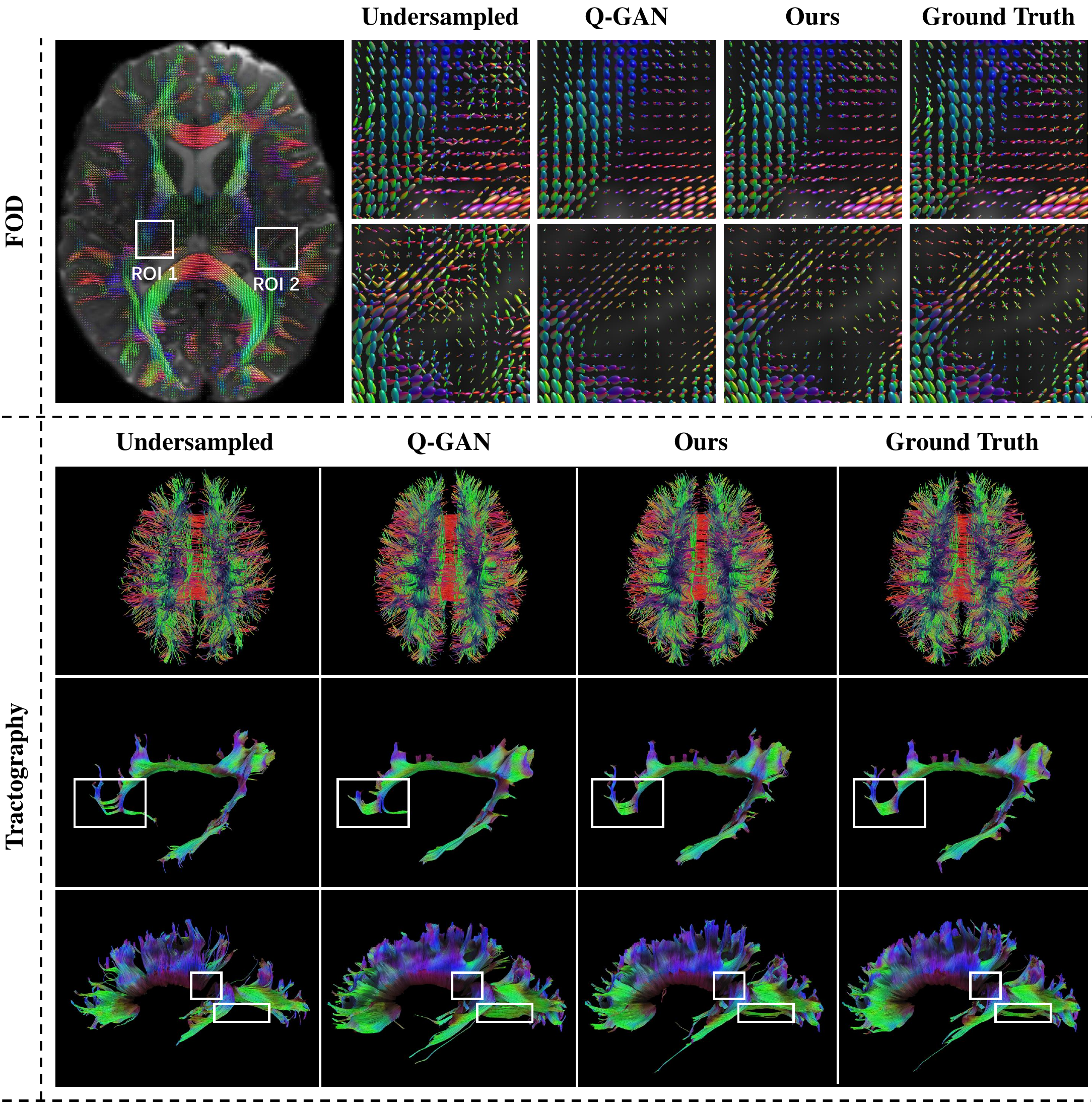}
	\end{minipage}
	\begin{minipage}{0.29\textwidth}
		\caption{Qualitative comparisons among different methods in fiber orientation distribution (FOD) and tractography. The top shows two magnified ROIs of FOD, and the bottom shows the tractography and two specific fiber tracts, i.e., Cingulum (L) and CorpusCallosum.} 
		\label{fig:fodtrack}
	\end{minipage}
	\vspace{-10pt}
\end{figure} 
To evaluate downstream performance beyond parameter fitting, we performed fiber orientation distribution (FOD) using MRtrix3 \cite{tournier2019mrtrix3} and FSL \cite{smith2004advances}, alongside tractography using DSI-Studio \cite{yeh2013deterministic}. Fig.~\ref{fig:fodtrack} shows that the fiber tracts generated from undersampled data exhibit the poorest quality, whereas those produced by Q-GAN better preserve continuity and integrity, and our results align more closely with the ground truth. These findings demonstrate that Q-CATN reliably and accurately represents underlying fiber structures in downstream analyses.

\section{Conclusion}\label{CD}
This study introduces a novel $q$-space guided collaborative attention translation network (Q-CATN) for MS-HARDI synthesis from flexible $q$-space sampling. Leveraging a collaborative attention mechanism, Q-CATN extracts complementary information from multiple modalities and modulates internal representations with flexible $q$-space conditions, addressing the limitations of fixed sampling strategies. Extensive experiments show that Q-CATN surpasses existing methods in synthesizing parameter maps and fiber tracts with fine-grained details, highlighting its potential for advancing diffusion applications. 
Notably, while recent diffusion models \cite{ho2020denoising} have shown impressive capabilities in generating images, Q-CATN was chosen for its superior computational efficiency and faster inference times, crucial for real-time clinical applications.\\

\noindent {\bf Acknowledgements.}
This research/project is supported by the STI 2030 -- Major Project (No. 2022ZD0209000), the National Research Foundation, Singapore, and the Agency for Science Technology and Research (A*STAR), Singapore, under its Prenatal/Early Childhood Grant (Grant No. H22P0M0007). Additional support is provided by the RGC GRF project (15201124) and the Hong Kong Global STEM Scholar scheme.\\

\noindent {\bf Disclosure of Interests.}
The authors have no competing interests to declare that are relevant to the content of this article.

\bibliographystyle{splncs04}
\bibliography{MICCAI2025}

\begin{thebibliography}{10}
\providecommand{\url}[1]{\texttt{#1}}
\providecommand{\urlprefix}{URL }
\providecommand{\doi}[1]{https://doi.org/#1}

\bibitem{chen2023deep}
Chen, G., Hong, Y., Huynh, K.M., Yap, P.T.: Deep learning prediction of
  diffusion mri data with microstructure-sensitive loss functions. Medical
  image analysis  \textbf{85},  102742 (2023)

\bibitem{cicimen2024image}
Cicimen, A.G., Tregidgo, H.F., Figini, M., Messaritaki, E., McNabb, C.B.,
  Palombo, M., Evans, C.J., Cercignani, M., Jones, D.K., Alexander, D.C.: Image
  quality transfer of diffusion mri guided by high-resolution structural mri.
  arXiv preprint arXiv:2408.03216  (2024)

\bibitem{daducci2015accelerated}
Daducci, A., Canales-Rodr{\'\i}guez, E.J., Zhang, H., Dyrby, T.B., Alexander,
  D.C., Thiran, J.P.: Accelerated microstructure imaging via convex
  optimization (amico) from diffusion mri data. Neuroimage  \textbf{105},
  32--44 (2015)

\bibitem{garyfallidis2014dipy}
Garyfallidis, E., Brett, M., Amirbekian, B., Rokem, A., Van Der~Walt, S.,
  Descoteaux, M., Nimmo-Smith, I., Contributors, D.: Dipy, a library for the
  analysis of diffusion mri data. Frontiers in neuroinformatics  \textbf{8}, ~8
  (2014)

\bibitem{gibbons2019simultaneous}
Gibbons, E.K., Hodgson, K.K., Chaudhari, A.S., Richards, L.G., Majersik, J.J.,
  Adluru, G., DiBella, E.V.: Simultaneous noddi and gfa parameter map
  generation from subsampled q-space imaging using deep learning. Magnetic
  resonance in medicine  \textbf{81}(4),  2399--2411 (2019)

\bibitem{golkov2016q}
Golkov, V., Dosovitskiy, A., Sperl, J.I., Menzel, M.I., Czisch, M., S{\"a}mann,
  P., Brox, T., Cremers, D.: Q-space deep learning: twelve-fold shorter and
  model-free diffusion mri scans. IEEE transactions on medical imaging
  \textbf{35}(5),  1344--1351 (2016)

\bibitem{gu2019generating}
Gu, X., Knutsson, H., Nilsson, M., Eklund, A.: Generating diffusion mri scalar
  maps from t1 weighted images using generative adversarial networks. In: Image
  Analysis: 21st Scandinavian Conference, SCIA 2019, Norrk{\"o}ping, Sweden,
  June 11--13, 2019, Proceedings 21. pp. 489--498. Springer (2019)

\bibitem{ho2020denoising}
Ho, J., Jain, A., Abbeel, P.: Denoising diffusion probabilistic models.
  Advances in neural information processing systems  \textbf{33},  6840--6851
  (2020)

\bibitem{isola2017image}
Isola, P., Zhu, J.Y., Zhou, T., Efros, A.A.: Image-to-image translation with
  conditional adversarial networks. In: Proceedings of the IEEE conference on
  computer vision and pattern recognition. pp. 1125--1134 (2017)

\bibitem{jensen2005diffusional}
Jensen, J.H., Helpern, J.A., Ramani, A., Lu, H., Kaczynski, K.: Diffusional
  kurtosis imaging: the quantification of non-gaussian water diffusion by means
  of magnetic resonance imaging. Magnetic Resonance in Medicine: An Official
  Journal of the International Society for Magnetic Resonance in Medicine
  \textbf{53}(6),  1432--1440 (2005)

\bibitem{karras2019style}
Karras, T., Laine, S., Aila, T.: A style-based generator architecture for
  generative adversarial networks. In: Proceedings of the IEEE/CVF conference
  on computer vision and pattern recognition. pp. 4401--4410 (2019)

\bibitem{mori2007introduction}
Mori, S.: Introduction to diffusion tensor imaging. Elsevier (2007)

\bibitem{park2021diffnet}
Park, J., Jung, W., Choi, E.J., Oh, S.H., Jang, J., Shin, D., An, H., Lee, J.:
  Diffnet: diffusion parameter mapping network generalized for input diffusion
  gradient schemes and b-value. IEEE Transactions on Medical Imaging
  \textbf{41}(2),  491--499 (2021)

\bibitem{ren2021q}
Ren, M., Kim, H., Dey, N., Gerig, G.: Q-space conditioned translation networks
  for directional synthesis of diffusion weighted images from multi-modal
  structural mri. In: Medical Image Computing and Computer Assisted
  Intervention--MICCAI 2021: 24th International Conference, Strasbourg, France,
  September 27--October 1, 2021, Proceedings, Part VII 24. pp. 530--540.
  Springer (2021)

\bibitem{smith2004advances}
Smith, S.M., Jenkinson, M., Woolrich, M.W., Beckmann, C.F., Behrens, T.E.,
  Johansen-Berg, H., Bannister, P.R., De~Luca, M., Drobnjak, I., Flitney, D.E.,
  et~al.: Advances in functional and structural mr image analysis and
  implementation as fsl. Neuroimage  \textbf{23},  S208--S219 (2004)

\bibitem{tian2020deepdti}
Tian, Q., Bilgic, B., Fan, Q., Liao, C., Ngamsombat, C., Hu, Y., Witzel, T.,
  Setsompop, K., Polimeni, J.R., Huang, S.Y.: Deepdti: High-fidelity
  six-direction diffusion tensor imaging using deep learning. NeuroImage
  \textbf{219},  117017 (2020)

\bibitem{tournier2019mrtrix3}
Tournier, J.D., Smith, R., Raffelt, D., Tabbara, R., Dhollander, T., Pietsch,
  M., Christiaens, D., Jeurissen, B., Yeh, C.H., Connelly, A.: Mrtrix3: A fast,
  flexible and open software framework for medical image processing and
  visualisation. Neuroimage  \textbf{202},  116137 (2019)

\bibitem{van2013wu}
Van~Essen, D.C., Smith, S.M., Barch, D.M., Behrens, T.E., Yacoub, E., Ugurbil,
  K., Consortium, W.M.H., et~al.: The wu-minn human connectome project: an
  overview. Neuroimage  \textbf{80},  62--79 (2013)

\bibitem{yan2016dpabi}
Yan, C.G., Wang, X.D., Zuo, X.N., Zang, Y.F.: Dpabi: data processing \&
  analysis for (resting-state) brain imaging. Neuroinformatics  \textbf{14},
  339--351 (2016)

\bibitem{yang2023towards}
Yang, J., Jiang, H., Tassew, T., Sun, P., Ma, J., Xia, Y., Yap, P.T., Chen, G.:
  Towards accurate microstructure estimation via 3d hybrid graph transformer.
  In: International Conference on Medical Image Computing and Computer-Assisted
  Intervention. pp. 25--34. Springer (2023)

\bibitem{ye2020improved}
Ye, C., Li, Y., Zeng, X.: An improved deep network for tissue microstructure
  estimation with uncertainty quantification. Medical image analysis
  \textbf{61},  101650 (2020)

\bibitem{yeh2013deterministic}
Yeh, F.C., Verstynen, T.D., Wang, Y., Fern{\'a}ndez-Miranda, J.C., Tseng,
  W.Y.I.: Deterministic diffusion fiber tracking improved by quantitative
  anisotropy. PloS one  \textbf{8}(11),  e80713 (2013)

\bibitem{yu2018multi}
Yu, X., Ying, Z., Li, T., Liu, S., Li, G.: Multi-mapping image-to-image
  translation with central biasing normalization. arXiv preprint
  arXiv:1806.10050  (2018)

\bibitem{zhang2012noddi}
Zhang, H., Schneider, T., Wheeler-Kingshott, C.A., Alexander, D.C.: Noddi:
  practical in vivo neurite orientation dispersion and density imaging of the
  human brain. Neuroimage  \textbf{61}(4),  1000--1016 (2012)

\bibitem{zheng2021spatially}
Zheng, C., Cham, T.J., Cai, J.: The spatially-correlative loss for various
  image translation tasks. In: Proceedings of the IEEE/CVF conference on
  computer vision and pattern recognition. pp. 16407--16417 (2021)

\bibitem{zong2024attention}
Zong, F., Zhu, Z., Zhang, J., Deng, X., Li, Z., Ye, C., Liu, Y.:
  Attention-based q-space deep learning generalized for accelerated diffusion
  magnetic resonance imaging. IEEE Journal of Biomedical and Health Informatics
   (2024)

\end{thebibliography}
\end{document}